\documentclass[twocolumn,preprintnumbers,amsmath,amssymb]{revtex4}

\usepackage{graphicx}
\usepackage{dcolumn}
\usepackage{bm}
\usepackage[final]{pdfpages}
\begin{document}
\title{Critical doping for the onset of a two-band superconducting ground state in SrTiO$_{3-\delta}$}
\author{Xiao Lin$^{1}$, German Bridoux$^{1}$, Adrien Gourgout$^{1}$, Gabriel Seyfarth$^{2}$, Steffen Kr\"{a}mer$^{2}$, Marc Nardone $^{3}$, Beno\^{\i}t Fauqu\'e$^{1}$ and Kamran Behnia$^{1}$\email{kamran.behnia@espci.fr}}
\affiliation{(1) Labotoire Physique et Etude de Mat\'{e}riaux-CNRS (ESPCI and UPMC), Paris, France\\
(2) Laboratoire National des Champs Magnétiques Intenses-CNRS (UJF, UPS and INSA), Grenoble , France
(3) Laboratoire National des Champs Magnétiques Intenses-CNRS (UJF, UPS and INSA), Toulouse , France}
\date{May 21, 2014}

\begin{abstract}
\textbf{In doped SrTiO$_{3}$ superconductivity persists down to an exceptionally low concentration of mobile electrons. This restricts the
relevant energy window and possible pairing scenarios. We present a study of quantum oscillations and superconducting transition temperature, $T_{c}$  as the carrier density is tuned from $10^{17}$ to $10^{20}$ $cm^{-3}$ and identify two critical doping levels corresponding to the filling thresholds of the upper bands. At the first critical doping, which separates the single-band and the two-band superconducting regimes in oxygen-deficient samples, the steady increase of T$_{c}$  with carrier concentration suddenly stops. Near this doping level, the energy dispersion in the lowest band displays a downward deviation from parabolic behavior. The results impose new constraints for microscopic pairing scenarios.}
\end{abstract}
\maketitle

Superconductiviy is induced in insulating SrTiO$_{3}$ by introducing n-type charge carriers through chemical doping\cite{schooley1} and survives
over three orders of magnitude of carrier concentration. The transition temperature, $T_{c}$, peaks to 0.45 K around a carrier density of
$n_{H}\sim10^{20}$ $cm^{-3}$\cite{schooley2}. A superconducting dome has also been detected in the metallic interfaces of SrTiO$_{3}$\cite{ohtomo} when the carrier density is modulated by a gate voltage bias\cite{caviglia}. In unconventional superconductors, such as high-T$_{c}$ cuprates, superconducting domes are often attributed to the proximity of a magnetic order or a Mott insulator. The recent discovery of superconducting dome in gate-tuned MoS$_{2}$\cite{ye} in absence of a competing order, however, highlights the limits of our current understanding of the interplay between carrier concentration and superconductivity and motivates a fresh reexamination of superconducting domes. In the specific case of SrTiO$_{3}$, superconductivity occurs in the vicinity of an aborted ferroelectric order\cite{muller} and survives deep inside the dilute metallic regime when the Fermi temperature becomes more than one order of magnitude lower than the Debye temperature\cite{lin}. This is a second puzzle in addition to the one raised by the drop in $T_{c}$ on the overdoped side. These two questions, raised at the opposite limits of the superconducting dome, remain unsettled.

\begin{figure}\resizebox{!}{0.7\textwidth}
{\includegraphics{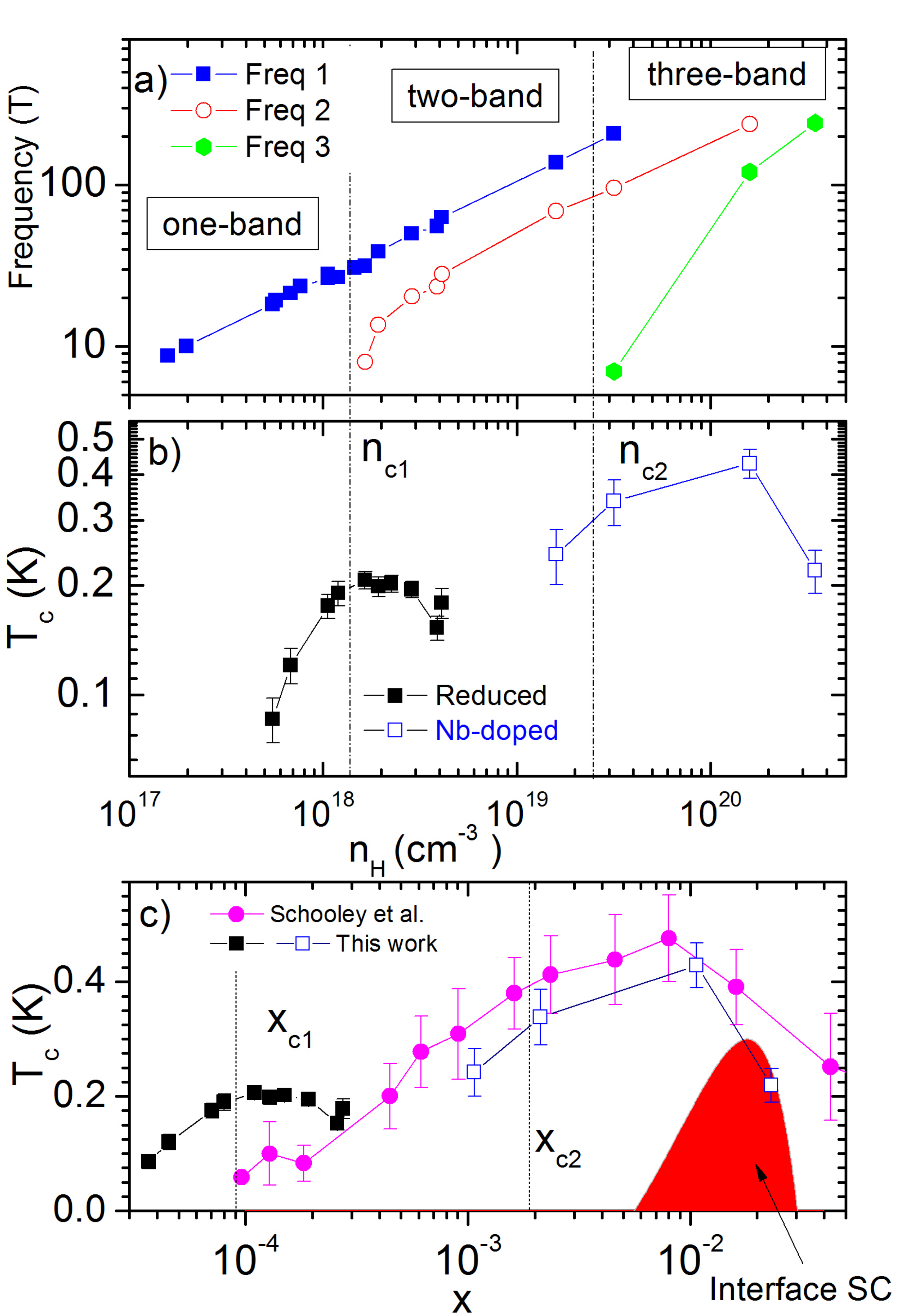}}
\caption{ a) Detected frequencies of quantum oscillations as a function of carrier concentration. At two critical doping levels, designated as n$_{c1}$ and n$_{c2}$, a new frequency emerges. At each critical doping a new band starts to be occupied. b) Superconduting resistive transition temperature (on a logarithmic axis) as a function of carrier concentration. Solid squares represent reduced samples (SrTiO$_{3-\delta}$) and open squares Nb-doped samples (SrTi$_{1-x}$Nb$_{x}$O$_{3}$ with x=0.02; 0.01;0.002 and 0.001). Error bars represent the width of transition. c) T$_{c}$ (on a linear axis) as a function of carrier per formula unit. Our data is compared with those reported by Schooley and co-workers\cite{schooley2}. The red shaded region shows the rough contours of superconductivity in STO/LAO interface\cite{caviglia,bell,richter}.}
\end{figure}

According to band calculations\cite{mattheiss,bistritzer,vandermarel}, doping SrTiO$_{3}$ with n-type carriers can fill three bands one after the other. Once the critical threshold for the occupation of a band is attained, a new Fermi surface concentric with the previous one emerges. Previous studies of quantum oscillations in bulk doped SrTiO$_{3}$\cite{gregory,uwe,lin,allen} have detected both multiple-frequency \cite{uwe,lin,allen} and single-frequency \cite{lin,allen} oscillations at different doping levels, but did not determine these critical doping levels. Moreover, according to tunneling experiments, doped SrTiO$_{3}$ beyond a carrier density of 10$^{19}$ $cm^{-3}$  is a multi-gap superconductor\cite{binnig}. The interplay between multi-band occupation in the normal state and multi-gap superconductivity has been a subject of recent theoretical attention\cite{fernandes}.

\begin{figure}\resizebox{!}{0.35\textwidth}
{\includegraphics{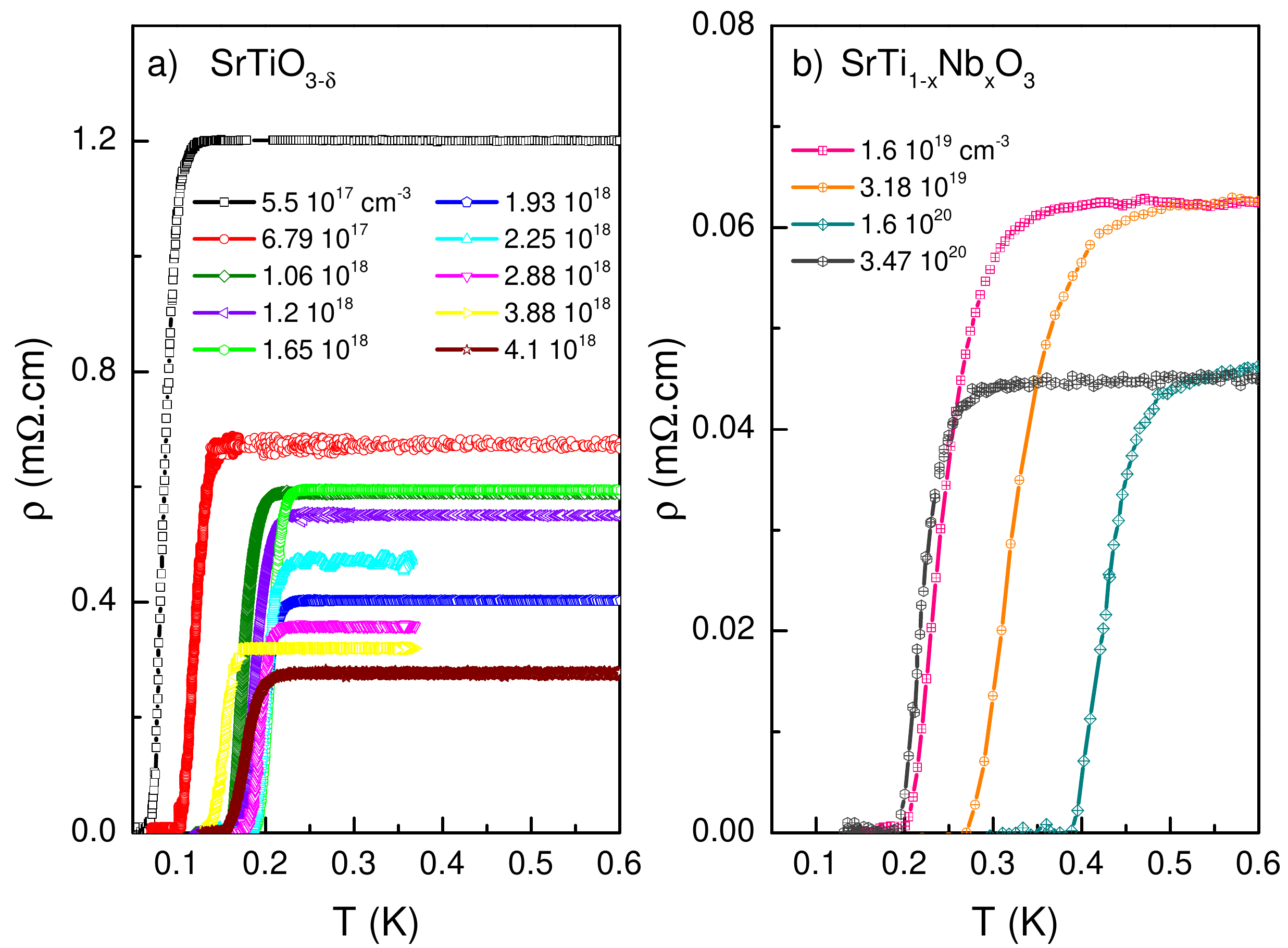}}
\caption{Low-temperature resistivity of  reduced (a) and Nb-doped (b) single crystals. Note the continuous evolution of resistive transitions in reduced samples. As the doping increases, the transition first shifts to higher temperatures and then remains more or less constant.}
\end{figure}

We present a systematic study of quantum oscillations and superconducting transition as a function of carrier concentration, $n_{H}$ extended
down to  10$^{17}$ $cm^{-3}$, two orders of magnitude below the range of the tunneling study\cite{binnig}, and find three new results. First of all, the
two critical dopings\cite{vandermarel}, $n_{c1}$ and $n_{c2}$ are identified and the magnitude of the cyclotron mass and the Fermi energy of each band are determined. Second, we find that $n_{c1}$, the threshold of occupation of the second band, separates two distinct regimes of superconductivity. Below $n_{c1}$, the  superconductor is single-band with a large $T_{c}/T_{F}$. Above $n_{c1}$, it becomes two-band with a $T_{c}$, which fails to keep the same pace with increasing T$_{F}$. Finally, we find that the lowest band presents a deviation from parabolic dispersion near n$_{c1}$. We conclude that the attractive interaction between electrons, remarkably strong when the chemical potential is near the bottom of this band, significantly weakens when it shifts upward and other bands are occupied. This feature, combined to the low cut-off frequency imposed by the small Fermi energy, drastically limits the choice of the bosonic mode exchanged by Cooper pairs.

\begin{figure*}\centering
\resizebox{!}{0.5\textwidth}
{\includegraphics{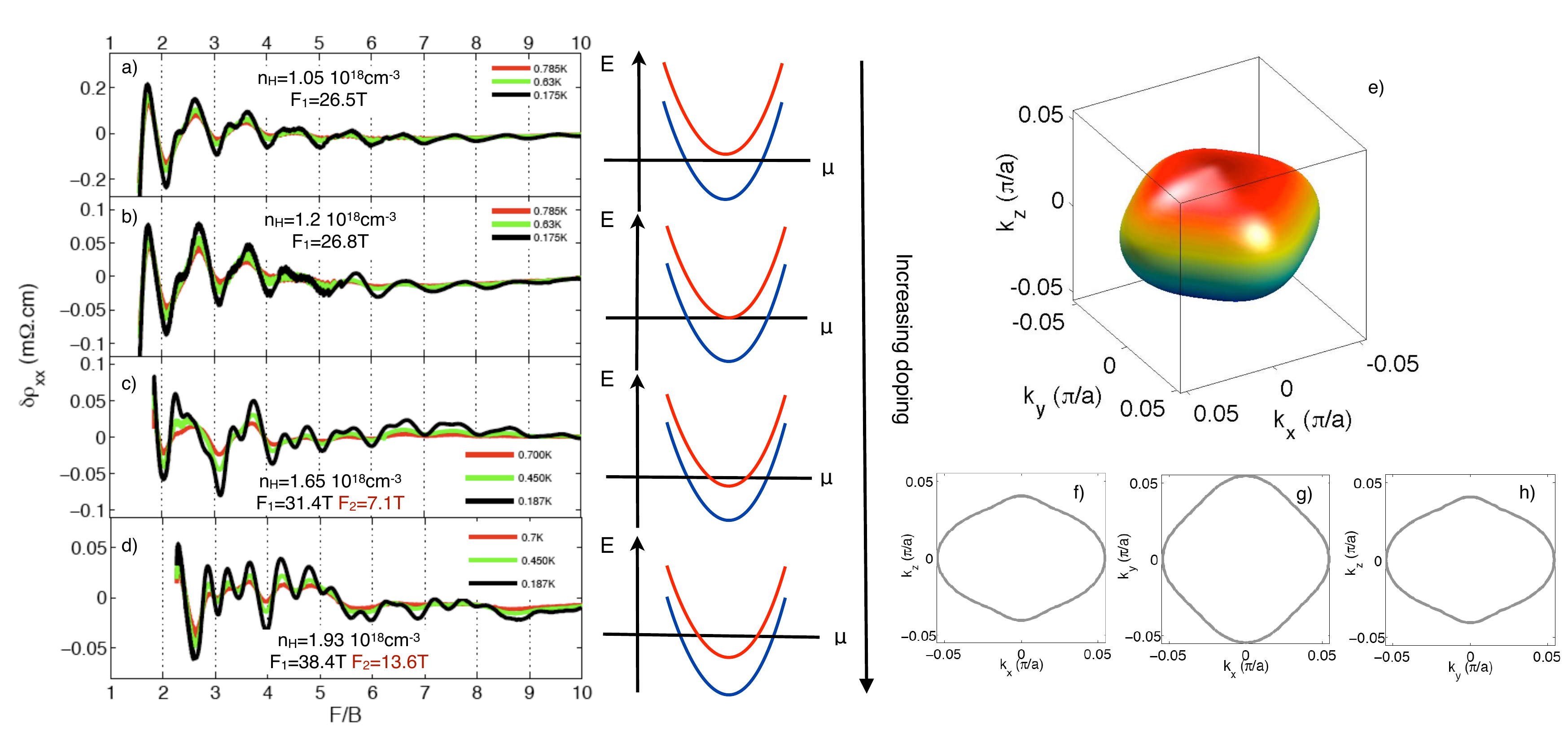}}
\caption{Left: A new frequency emerges in the Shubnikov-de Haas data in the vicinity of n$_{c1}$. When the carrier density is below n$_{c1}$, quantum oscillations display a single periodicity with split peaks and each peak is more prominent
than its preceding neighbor(panel a). With increasing carrier density, a second frequency emerges on top of the first one (panels b to d). Data is plotted as function of the inverse of magnetic field normalized by the principal frequency, $F_{1}$, in order to simplify comparison.  The Fermi surface of the lower band according to ref.\cite{allen} is shown in panel e and its projection along the three high-symmetry axes in panels f to g. In multi-domain crystals, the two possible Fermi cross-sections give rise to split peaks in quantum oscillations.}
\end{figure*}

The top panel in Fig. 1 shows the variation of the frequency of the quantum oscillations with doping in eighteen different samples of n-doped SrTiO$_{3}$ [See the supplement for an extensive discussion of the samples]. The samples studied were either  obtained by annealing stoichiometric SrTiO$_{3}$ in vacuum\cite{spinelli,lin} or commercial niobium-doped samples. In most cases, what was measured was the Shubnikov-de Haas(SdH) effect (quantum oscillations of the magnetoresistance). In seven samples, the Nernst effect was also measured and found to display giant oscillations with a frequency identical to the SdH effect as reported previously\cite{lin}. As seen in the figure, the main frequency smoothly evolves with doping. Moreover, at two critical doping levels new frequencies emerge. Panel b shows the evolution of the superconducting transition temperature of the same samples. The  most underdoped samples ($n_{H} < 4\times10^{17}cm^{-3}$) did not show a superconducting transition down to 60 mK. Nevertheless, they presented a sharp Fermi surface, indicating that superconductivity is preceded by (or concomitant with) the intersection of the chemical potential and the bottom of the conduction band. The most striking feature of the figure is a clear change in the slope of T$_{c}$(n) in the vicinity of  n$_{c1}$. This is the first new result of this study.

The bottom panel in Fig. 1 compares our data with the early work reported by Schooley\emph{et al.}\cite{schooley2}. For each sample, the error bar represents the width of resistive transition, i.e. the interval between two temperatures corresponding to $0.1$ and $0.9$ drops in resistivity (See our data in Fig. 2). While the two sets of data match roughly with each other, the structure in the dilute limit is far more clear in our data.  The panel shows also the interface dome, restricted to a narrow window far above the concentration range scrutinized by this work. The evolution of resistive superconducting transitions with carrier concentration across n$_{c1}$  is visible in Fig. 2.

The SdH data leading to an unambiguous identification of  n$_{c1}$ is presented in Fig. 3. The oscillating component of magnetoresistance, obtained after the subtraction of a smooth background (obtained ny a polynomial fit), is plotted as a function of the inverse of magnetic field times the main frequency. The sample with the lowest carrier concentration shows a single set of oscillations with split peaks. We attribute this splitting to the presence of tetragonal domains in our crystals. Recently, Allen\emph{ et al.}\cite{allen}, by performing an angle-dependent Shubnikov-de Haas study, found that the Fermi surface of the lower band  is larger along $k_{z}$ than along either $k_{x}$ or $k_{y}$. In our study, the magnetic field is applied along the cubic [100] axis. Below 105 K, in the tetragonal phase, our samples can host three domains corresponding to the three possible orientations for the tetragonal $z$-axis.  As illustrated in Fig. 3b, there would be two possible magnitudes for the cross section of the Fermi surface, when the magnetic field is along the nominal [100] axis. In this case, the main frequency corresponds to the smaller cross section, which is two times more likely to occur than the larger one. As discussed in detail in the supplement, this interpretation is in very good agreement with the known topology of the Fermi surface.

As seen in Fig. 3a, at the concentration of 1.2 10$^{18}$ cm$^{-3}$ beating emerges  and indicates the presence of a second set of oscillations with a significantly longer period. In samples with $n_{H} >n_{c1}=1.2\times10^{18}$ $cm^{-3}$, the presence of this second frequency becomes clear and then it rapidly grows with increasing doping. The frequencies were quantified both by a FFT analysis of our data [See the supplement] and direct identification of the two periods in the data.

\begin{figure*}\centering
\resizebox{!}{0.65\textwidth}
{\includegraphics{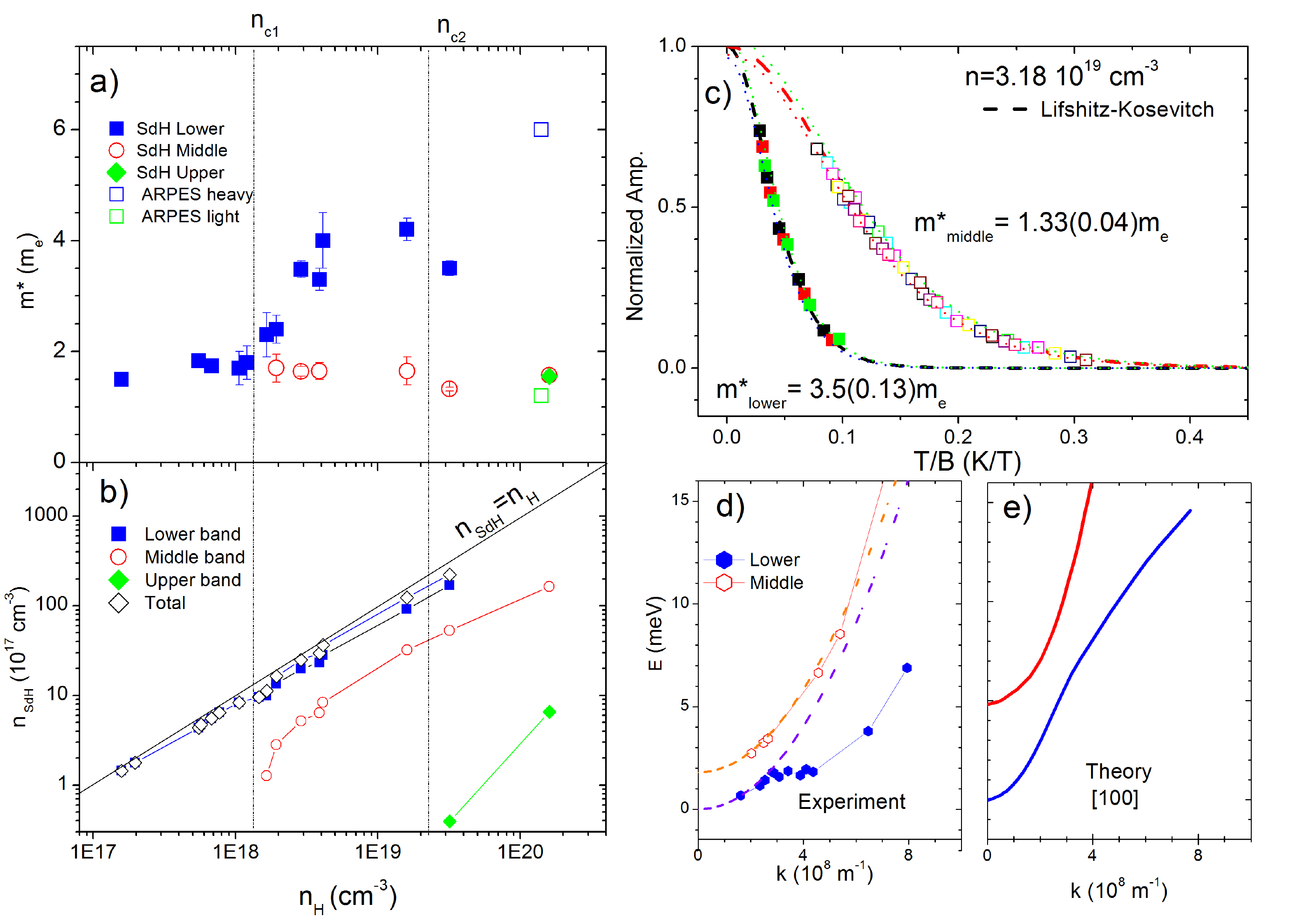}}
\caption{ a) The effective mass in each band as a function of carrier concentration. The masses reported by an ARPES study on a bulk sample at $ n=1.4 \times10^{20}$ $cm^{-3}$\cite{chang} is also shown for comparison. b) Carrier concentration and its distribution among bands according to the detected frequency and assuming spherical Fermi surfaces. The total carrier density falls just below the Hall carrier density. Above $1.6\times10^{20}$ $cm^{-3}$, the lower band becomes invisible because of its reduced mobility. c) Lifshitz-Kosevitch analysis of the oscillations revealing two distinct masses for the lower and  middle bands. Each color corresponds to a different oscillation. d) The Fermi energy, defined as $\frac{\hbar^{2}k_{F}^{2}}{2m_{c}}$ in the two lower bands plotted as a function of corresponding $k_F$. The middle band is shifted upward by the energy gap between the two bands. e) Theoretical energy dispersion near the bottom of the bands according to ref.\cite{vandermarel}. }
\end{figure*}

Analysis of quantum oscillations using the Lifshitz-Kosevitch formalism yields the magnitude of the cyclotron mass. The results of the analysis
are shown in the upper panel of Fig. 4. We find that below n$_{c1}$, the effective cyclotron mass in the lower band, $m_{1}$ keeps a steady magnitude of 1.5 to 1.8m$_{e}$, as previously reported \cite{lin}. At higher doping levels, $m_{1}$ becomes as large as 4$m_{e}$. In this doping range, the middle
band hosts significantly lighter carriers ($m_{2}\sim 1.3m_{e}$).

Comparing the total density of carriers distributed in the three bands and the carrier concentration obtained by measuring the Hall number, one can check the consistency of the emerging overall picture. Assuming three spherical Fermi surfaces leads to Fig. 4b. As seen in the figure, up to $n_{c2}$ the two concentrations remain close to each other. The ratio of the two remains lower than unity, which is an expected consequence of the contrast between the assumed sphericity of the Fermi surface and its moderate anisotropy according to both experiment\cite{uwe,allen} and theory\cite{mattheiss,vandermarel}. As discussed in the supplement, the measured anisotropy of the Fermi surface below n$_{c1}$ \cite{allen} gives a satisfactory account of $\frac{n_{SdH}}{n_{H}}$ in our data. At the carrier density of 1.6 10$^{20}cm^{-3}$, three bands are filled, but experiment detects only two frequencies (See Fig. 1a). It is straightforward to identify both the missing band and the reason of its absence. Only in the two upper bands the mobility is high enough for detectable quantum oscillations in a magnetic field of 17 T. The total density of carriers concentrated in these two visible bands is well below the total carrier concentration (See Fig. 4b). Both these features point to the lower band (which contains most of carriers, but those with less mobility) as the one which becomes invisible at this doping level.

Now, we can address a central issue in the description of the band structure in this system. According to the picture first sketched by Mattheiss\cite{mattheiss},  the t$_{2g}$ orbitals of titanium atoms give rise to a threefold degenerate band\cite{bistritzer,vandermarel}. The degeneracy is lifted by a combination of spin-orbit coupling and tetragonal distortion. Recently, van der Marel and co-workers\cite{vandermarel}  pointed out that the energy gap between the bands and the peculiarity of their dispersion crucially depends on the choice of parameters. Our experiment finds that the effective mass in the lower band begins to increase as soon as the average radius of the Fermi surface exceeds $3 \times 10^{8}$ $m^{-1}$. As illustrated in panels $d$ and $e$ of Fig. 4, this feature is in good agreement with the choice of parameters in ref.\cite{vandermarel}, which leads to a non-parabolic dispersion for the lower band in absence of interaction. On the other hand the amplitude of the cyclotron mass and the size of the gaps differs significantly between non-interacting theory and experiment. The experimentally determined mass at the bottom of the band ($1.5 m_{e}$) is twice as large as the theoretical expectation\cite{vandermarel}. As a consequence of mass renormalization, commonly attributed to electron-phonon interaction\cite{vanmechelen}, the gap between the lower and middle bands (the Fermi energy of the lower band at $n=n_{c1}$) is 1.8 meV, significantly lower than the  expected value of 4.3 meV in the non-interacting band picture\cite{vandermarel}.

The main outcome of this study is to identify a critical doping, n$_{c1}$, below which a single band is occupied and T$_{c}$ rapidly rises with doping. In this regime, the $T_{c}/T_{F}$ ratio is as large as 0.01, comparable to many unconventional superconductors\cite{lin}. This process is interrupted when carrier concentration exceeds $n_{c1}$. This highlights the specificity of multi-band superconductivity in SrTiO$_{3}$, the subject of a recent theoretical work\cite{fernandes}. Motivated by the experimental reports on variation of T$_{c}$ near n$_{c2}$\cite{binnig}, Fernandes and co-workers considered the variation of superconducting transition temperature across a critical doping for different combinations of positive interband ($\lambda_{ij}$) and intraband ($\lambda_{ii}$) superconducting coupling parameters and found that filling an upper band (labeled 2) in addition to a lower band (labeled 1) would enhance T$_{c}$. The larger the ratios of  $\lambda_{12}/ \lambda_{11}$ and/or $\lambda_{22}/\lambda_{11}$, the more drastic is the expected enhancement. This contrasts with our experimental observation of a \emph{drop} in $T_{c}(n)$ near n$_{c1}$.

This discrepancy calls for a theoretical reexamination extended to  negative (i.e. repulsive) $\lambda_{22}$ and $\lambda_{12}$. More importantly, in this analysis, $\lambda_{ii}$ has been taken to be a constant $V_{i}$ (representing overall attractive interaction between electrons) times $N_{i}(0)$, the density of states at the Fermi energy at band $i$. This assumption can only generate a monotonously increasing critical temperature, since the density of states unavoidably increases with carrier concentration. The existence of domes implies that this approximation does not hold and points to a non-trivial evolution of $V_{1}$ as the outer Fermi surface expands.

Below n$_{c1}$,  the normal state of superconducting SrTiO$_{3}$, is a metal with a Fermi energy  as low as 1 meV and thus the bosons exchanged between pairing electrons cannot be typical phonons with a Debye energy more than one order of magnitude larger\cite{lin}. One candidate to replace them at small wave-vectors is a plasmonic mode suggested by Takada\cite{takada}. Another one, invoked by Appel\cite{appel}, is the soft mode observed below the 105K structural transition\cite{shirane}. Our results frame the challenge faced by such propositions in a multi-gap context. The attractive interaction should couple electronic states more efficiently below n$_{c1}$, i.e. when the Fermi-surface is single component and the dispersion still parabolic.

In spite of its notorious inability to predict new superconductors\cite{hirsch}, the BCS theory has been successfully employed to give an account of unexpectedly discovered superconductors such as MgB$_{2}$\cite{choi}. In the case of SrTiO$_{3}$, the challenge to theory is framed with exceptional
sharpness and few \emph{ad hoc} parameters. The combination of a simple Fermi surface topology, a very narrow energy window and a wide superconducting dome with fine details provides an exceptional opportunity to test the expected link between the critical temperature, the electronic density of states and the effective electron-electron interaction.

We thank A. Balatsky, R. Fernandes, D. van der Marel, I. Mazin and J. -M. Triscone for discussions. This work is supported  by Agence Nationale de la Recherche as part of QUANTHERM and SUPERFIELD projects and by EuroMagNET II under the EU contract number 228043. K. B. acknowledges the hospitality of Aspen Physics center.

\newpage
\appendix

\begin{table*}\centering
 \renewcommand{\thetable}{S1}
 \caption{n-doped SrTiO$_{3}$ single crystals studied in this work.}
 \begin{ruledtabular}
\begin{tabular}{c|c|c|c|ccc|ccc}
   Sample   & $n_H$    & $\mu_H$                 & $T_{c}$ & $F_1$ & $F_2$ & $F_3$& $m^*_1$ & $m^*_2$  & $m^*_3$  \\
 &  cm$^{-3}$       & cm$^2$.V$^{-1}$.s$^{-1}$&  mK         &    T     & T        & T       & $m_e$   & $m_e$    & $m_e$     \\
\hline

SrTiO$_{3-\delta}$ & 1.58 10$^{17}$ & 10395       &  a         &  8.7     &--        &--       & 1.5(0.1)&--        &--  \\

SrTiO$_{3-\delta}$ & 5.5 10$^{17} $ &  8941        &  86         &  18.2    &--         &--       & 1.83(0.07)&--        &--  \\

SrTiO$_{3-\delta}$ & 5.7 10$^{17} $ &  N. M.       &  N. M.         &  19.2    &--         &--       & N. D. &--        &--  \\

SrTiO$_{3-\delta}$ & 6.79 10$^{17}$   & 13670      &  120.5         &  21.3    &--      &--       & 1.74(0.1)&--        &--  \\

SrTiO$_{3-\delta}$ & 7.67 10$^{17}$   & 7409      &  N. M.         &  23.5    &--      &--       & N. D. &--        &--  \\

SrTiO$_{3-\delta}$ & 1.06 10$^{18}$ & 9949        &  175          &  26.5    &--         &--       & 1.7(0.3)&--        &--  \\

SrTiO$_{3-\delta}$ & 1.2 10$^{18}$ & 9385          &  190        &  26.8    &--         &--       & 1.8(0.3)&--        &--  \\

SrTiO$_{3-\delta}$ & 1.65 10$^{18}$ &  6366        &  206        &  31.4    &7.1         &--       & 2.3(0.4)&N. D.       &--  \\

SrTiO$_{3-\delta}$ & 1.93 10$^{18}$ &  8060        &  198        &  38.4   &13.6        &--       & 2.4(0.25)&1.7(0.25)       &--  \\

SrTiO$_{3-\delta}$ & 2.25 10$^{18}$ &  5868        &  202        &  N. M.   & N. M.        &--       & N. M. & N. M.       &--  \\

SrTiO$_{3-\delta}$ & 2.88 10$^{18}$ &  6078        &  194        &  49.8  &20.4        &--       & 3.48(0.15)&1.64(0.09)    &--  \\

SrTiO$_{3-\delta}$ & 3.88 10$^{18}$ &  5066        &  153        &  55.5  &23.4        &--       & 3.3(0.2)&1.65(0.15)    &--  \\

SrTiO$_{3-\delta}$ & 4.11 10$^{18}$ & 5421         &  178.5      &  63    &28 & --         & 4(0.5) & N. D.    &--  \\
\hline
SrTi$_{0.999}$Nb$_{0.001}$O$_3$ & 1.6 10$^{19}$   & 6290         &  242       &  138   &69       & --           & 4.2(0.2) & 1.65(0.25)   &--  \\

SrTi$_{0.998}$Nb$_{0.002}$O$_3$ & 3.18 10$^{19}$ &  3121        &  338.6      &  208   &96       &7          & 3.5(0.13) & 1.33(0.04)   &N. D. \\

SrTi$_{0.99}$Nb$_{0.01}$O$_3$ & 1.6 10$^{20}$ &  870            &  429      &  N. D.  &204       &116          & N. D. & 1.57(0.08)   & 1.55(0.08)  \\

SrTi$_{0.98}$Nb$_{0.02}$O$_3$ & 3.47 10$^{20}$ &  406           &  219     &  N. D.   &N. D.       &243          & N. D. & N. D.   & N. D.  \\
\hline
SrTi$_{0.9996}$Nb$_{0.0004}$O$_{3-\delta}$[b] & 1.97 10$^{17}$ &   10600   &  N. M.     &  10   &--    &   --      & N. D. & --   & --  \\

SrTi$_{0.9996}$Nb$_{0.0004}$O$_{3-\delta}$[b] & 1.06 10$^{18}$ &   18533   &  a     &  28   &--    &   --      & 1.9 (0.2) & --   & --  \\

SrTi$_{0.9996}$Nb$_{0.0004}$O$_{3-\delta}$[b] & 1.46 10$^{18}$ &  14444   &  a     &  30.7   &--      & --        & 1.9 (0.2) & --   &-- \\

SrTi$_{0.9996}$Nb$_{0.0004}$O$_{3-\delta}$[b] & 2.5 10$^{18}$ &  14229   &  130     &  N. M.   &N. M.      & --       & N. M. & N. M.   & --  \\
\end{tabular}
\end{ruledtabular}
\begin{tabular}{ccc}
\multicolumn{3}{l}{a. Superconductivity was not detected down to 60 mK}\\
\multicolumn{3}{l}{b. Nominal content (See text).}\\
\multicolumn{3}{l}{c. N. D. refers to quantities which could not be determined.}\\
\multicolumn{3}{l}{d. N. M. refers to quantities which were not measured.}\\
\multicolumn{3}{l}{e. Irrelevant quantities are specified by --.}
\end{tabular}
\label{TabI}
\end{table*}

\section{\label{sec:level1}Samples}
Table S1 lists the properties of the samples used in this study. The table indicates Hall carrier concentration (n$_H$),  Hall mobility ($\mu_H$), the superconducting critical temperature (T$_{c}$, defined as the temperature at which the resistivity attains half of its normal-state value), oscillation frequencies ($F_{i}$) and cyclotron mass ($m_{i}^*$) for each sample. Oscillation frequencies and cyclotron mass were extracted from the quantum oscillations of magnetoresistance or the Nernst coefficients.

Three different types of electron-doped SrTiO$_{3}$ single crystals were utilized in this study. The first category  were  oxygen deficient SrTiO$_{3-\delta}$ obtained by annealing stoichiometric SrTiO$_{3}$ samples in a temperature range of 700-1000 $^{0}$C in a vacuum of 10$^{-6}$ mbar for 1 to 2 hours\cite{spinelli,XLin}. The concentration of oxygen vacancies was controlled by finely tuning the annealing temperature. A second category of samples were commercially- bought Nb-doped SrTiO$_{3}$ single crystals. Four samples with well-defined Nb content (SrTi$_{1-x}$Nb$_{x}$O$_3$; with x=0.001;0.002;0.01;0.2) were studied. In these four samples, the carrier concentration determined from the Hall coefficient matched the expected carrier concentration assuming that substitution of a titanium atom by a niobium atom introduces a single mobile electron.

A third category of the samples were made by reducing a Nb-doped sample with  a nominal Nb concentration of x= 0.0004. Before reduction, the carrier concentration of this nominally SrTi$_{0.9996}$Nb$_{0.0004}$O$_3$ single crystal was orders of magnitude lower than the expected concentration of $8 \times10^{18}cm^{-3}$. Preparing reduced samples cut from this anomalously doped sample produced unexpected samples, which showed high mobility and displayed quantum oscillations of expected frequency, but their transition temperature was anomalously low.

In our analysis, we have focused only on the first two types of samples in which the origin of mobile electrons introduced by chemical substitution is clearly identified. Carriers are introduced either by removing an oxygen atom, or substituting a titanium atom by a niobium atom.

\begin{figure}
\renewcommand{\thefigure}{S1}
\resizebox{!}{0.38\textwidth}
{\includegraphics{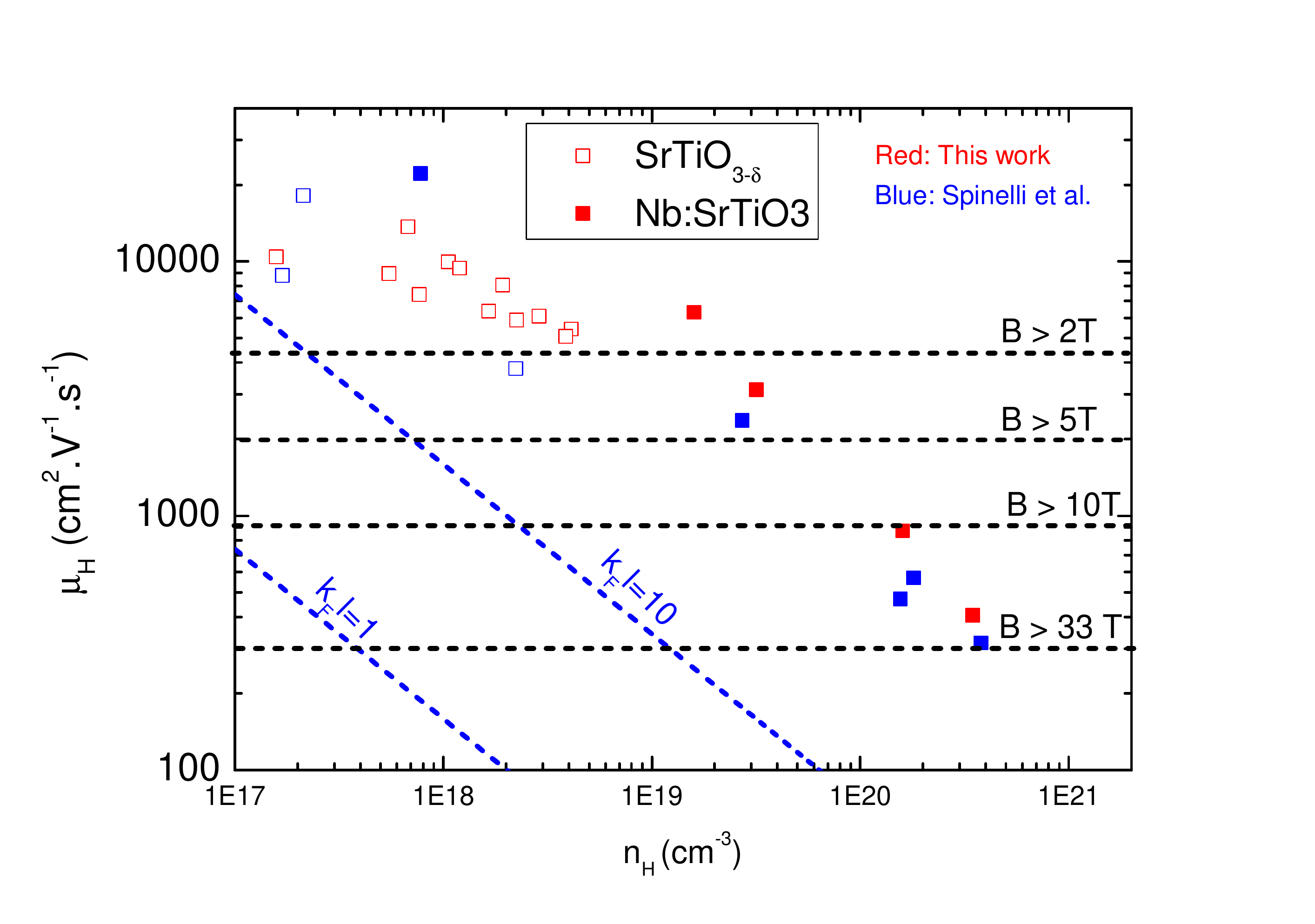}}
\caption{Mobility of Nb-doped and reduced SrTiO$_{3}$ samples as a function of carrier concentration. The horizontal black lines represent the mobility required to observe quantum oscillations in a given magnetic field. The blue lines represent the mobility threshold to avoid a metal-insulator transition. }
\end{figure}

\section{\label{sec:level1}Shubnikov -de Haas oscillations}

\begin{figure*}\centering
\renewcommand{\thefigure}{S2}
\resizebox{!}{0.6\textwidth}{\includegraphics{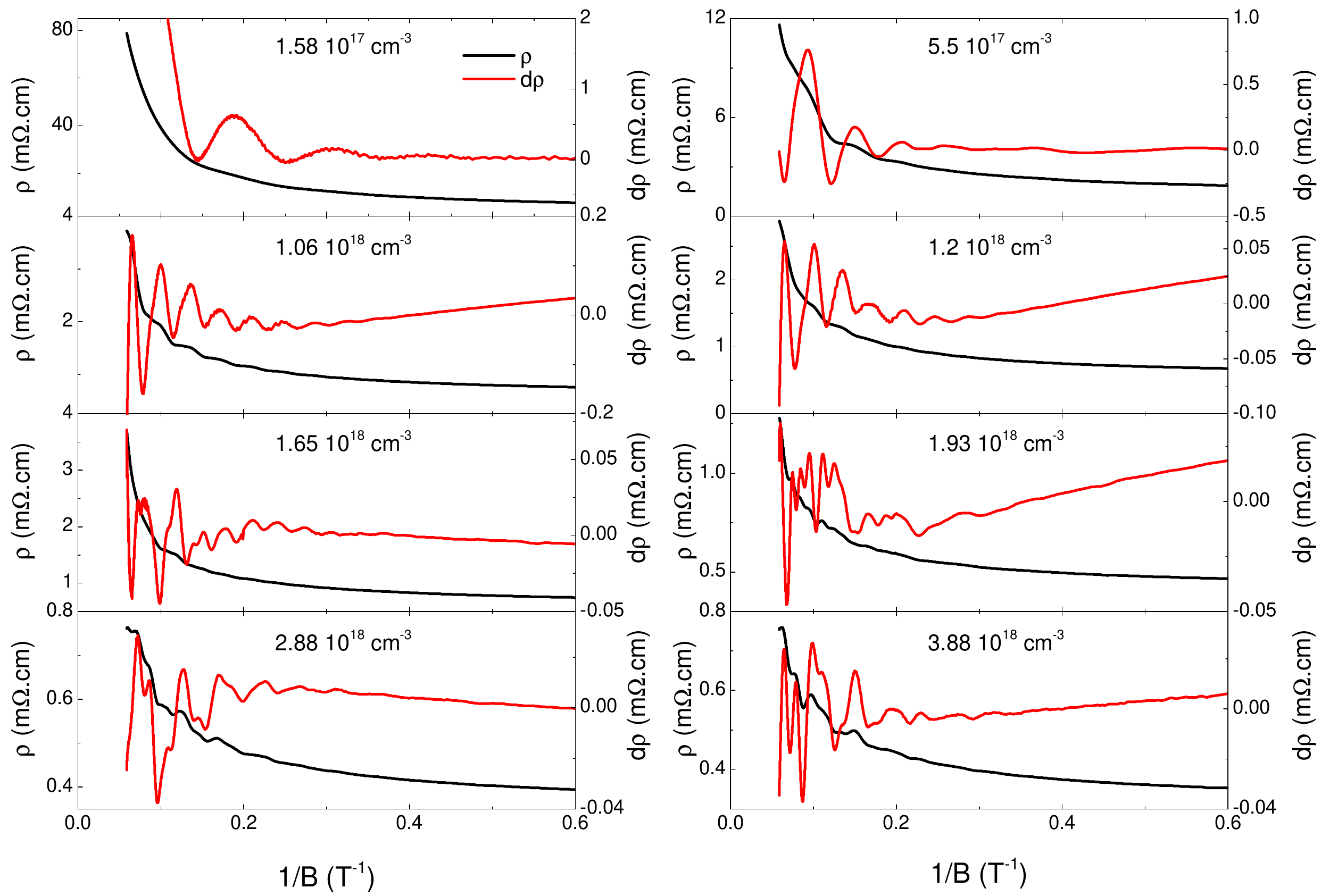}}
\caption{The raw magneto-resistance ($\rho$) and residual resistance (d$\rho$) as the function of the inverse field (1/B) at T = 550 mK and B // [0,0,1]. d$\rho$ is attained from $\rho$ by subtracting the non-oscillating high temperature resistance curve.}
\end{figure*}

Fig. S1 compares the mobility in our samples with those reported by Spinelli \emph{et al.} \cite{spinelli}.  To observe quantum oscillations electrons are required to be mobile enough to avoid a collision during a cyclotron period (implying $\mu >B^{-1}$ ) and also to avoid localization ($k_{F}l >1$, implying $\mu>e\hbar^{-1}k_{F}^{-2}$). As seen in the figure, all samples studied here have mobilities high enough to meet these two requirements.

Fig. S2 shows the magnetoresistance ($\rho$) and its oscillation component(d$\rho$) as a function of B$^{-1}$  at T=0.55 K for different samples with different carrier concentrations.  As seen in the figure, the amplitude of both the total magneto-resistance and its oscillating component decreases as the carrier concentration increases. The oscillation structure is simple and shows single periodicity in the upper panel. With the increase in doping, the oscillations become more complex.

Fig. S3 shows the normalized  amplitudes of FFT in different samples. As seen clearly, there is a single  frequency below n = 1.65 10$^{18}$ cm$^{-3}$, implying that only the lowest band is occupied. A second frequency of 8T emerges at n = 1.65 10$^{18}$ cm$^{-3}$, implying that the middle band starts to be filled. The frequencies become larger by electron doping, indicating an expanding Fermi surface.

\begin{figure*}\centering
\renewcommand{\thefigure}{S3}
\resizebox{!}{0.35\textwidth}
{\includegraphics{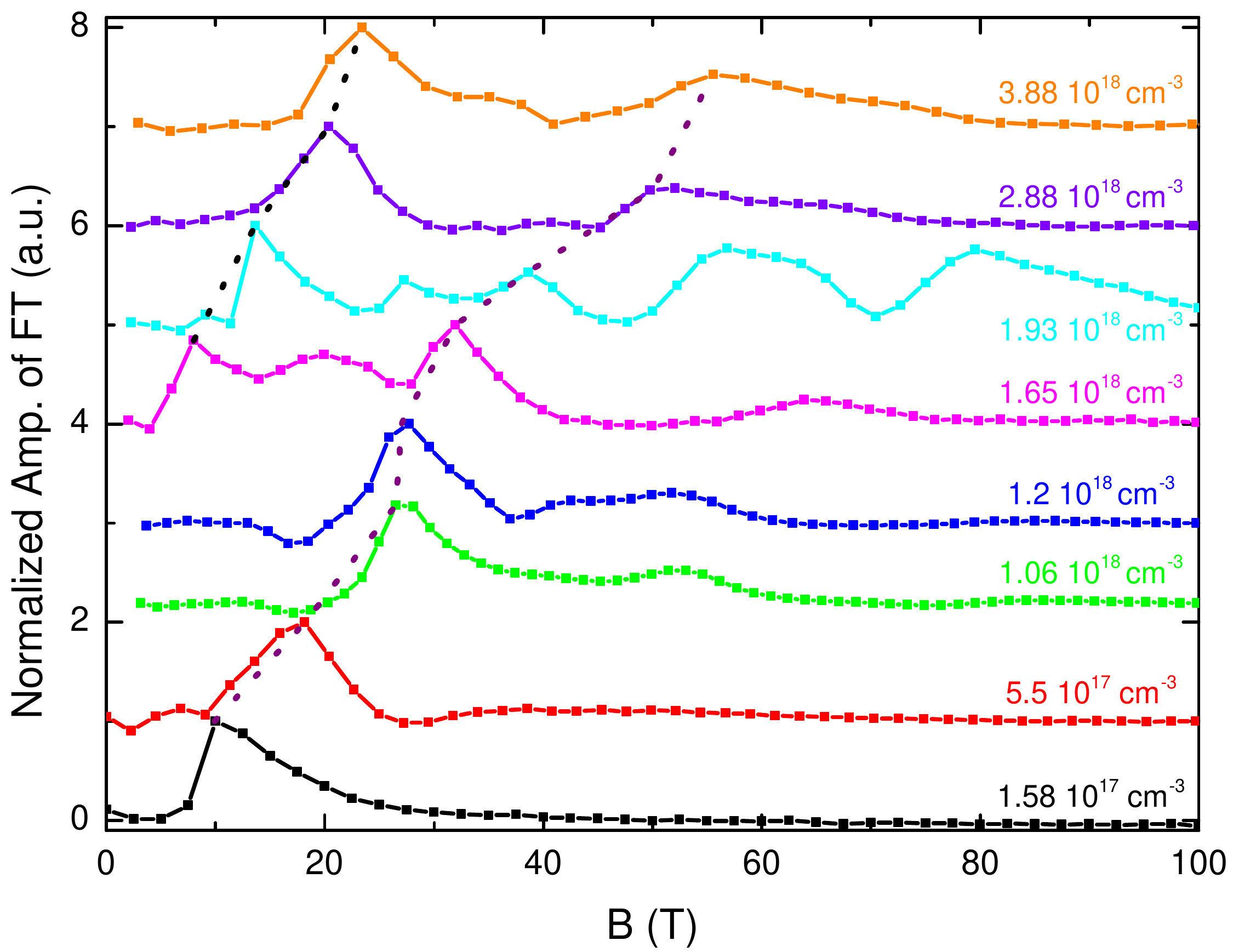}}
\caption{ The normalized Fast Fourier transform (FFT) of the oscillating component at T = 550mK for different samples. The dotted lines show the evolution of frequency of each band as the carrier concentration increases.}
\end{figure*}

\section{\label{sec:level1}Fermi surface distortion and the split peaks in multi-domain samples}

Even at very low doping, near the bottom of the lower band, the Fermi surface of $n-$doped SrTiO$_{3}$ is not a perfect sphere\cite{uwe}. Working on single-domain La-doped crystals of SrTiO$_{3}$, Allen and co-workers quantified its departure from sphericity\cite{allen}. They found that the Fermi surface is squeezed along the c-axis such that the area of its projection on the ($k_{x},k_{y}$) plane is larger than its projection on the ($k_{x},k_{z}$) or the ($k_{y},k_{z}$) planes.

In a multi-domain crystal, the tetragonal distortion can occur along any of the three principal axes. Since the projection of the Fermi surface on the ($k_{x},k_{z}$) and ($k_{y},k_{z}$) planes are identical, with equal populations of the three domains, the main detected frequency would correspond to the projection of the Fermi surface along this direction. In this case, an additional frequency is expected corresponding to the projection of the Fermi surface on the ($k_{x},k_{y}$) plane.  This gives a natural explanation for the presence of split peaks in our data. Fig. S4a compares the evolution of main frequencies according to our data and those reported  by Allen and co-workers\cite{allen}. As seen in the figure, our main frequency is close to the one they detect when the field is along the smaller cross section of the Fermi surface. This confirms the interpretation proposed here.

\begin{figure*}\centering
\renewcommand{\thefigure}{S4}
\resizebox{!}{0.4\textwidth}
{\includegraphics{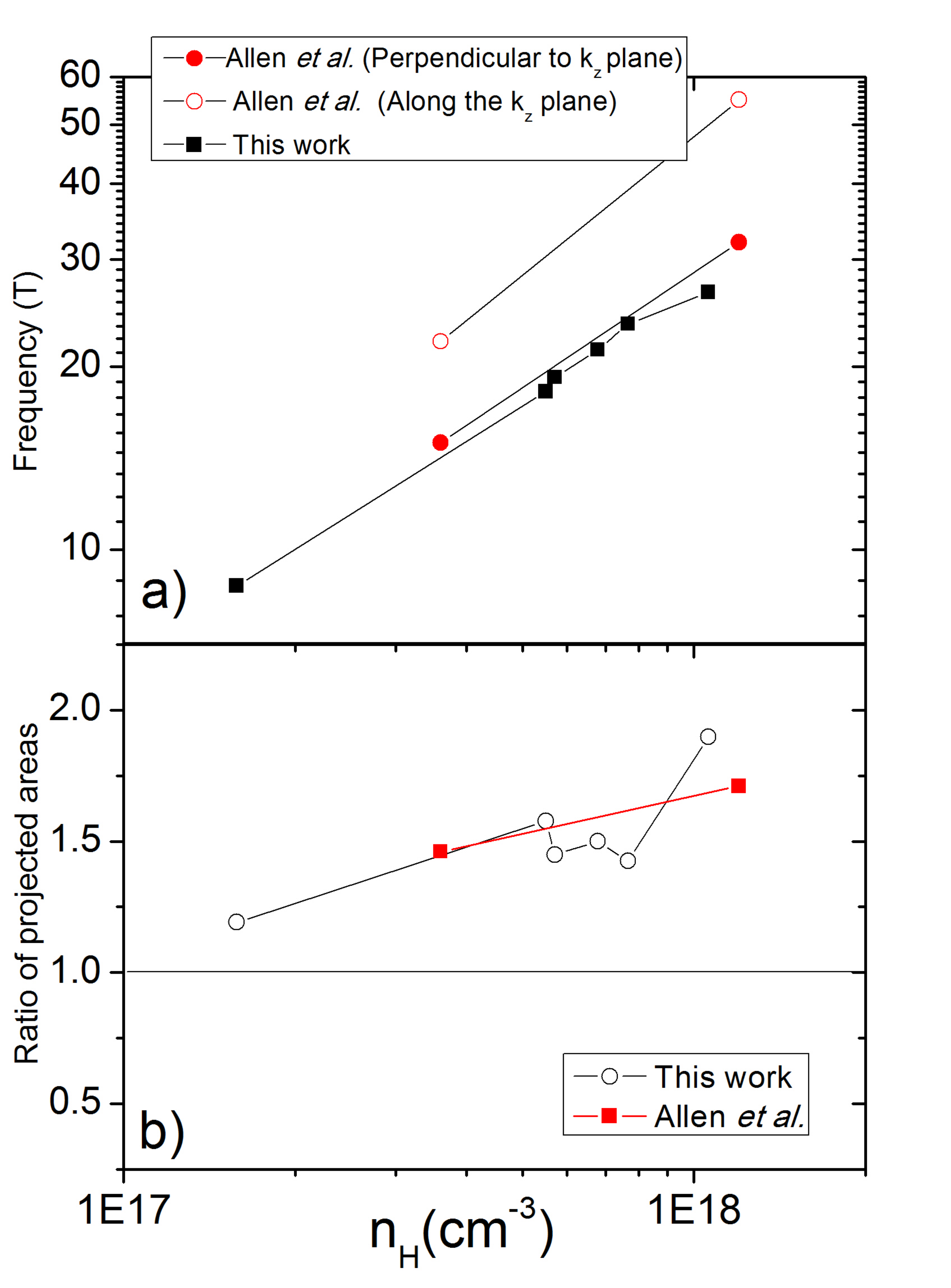}}
\caption{ a) SdH frequencies as a function of carrier density in this work compared to the study by Allen \emph{et al.}.b) The magnitude of tetragonal distortion according to the two studies. }
\end{figure*}

A second quantitative confirmation comes from  the magnitude of the tetragonal distortion determined by Allen and co-workers\cite{allen}, who directly measured the frequency of oscillations for a field perpendicular to the large face of the Fermi surface (i.e. along k$_z$)  and perpendicular to it. The ratio of these frequencies was found to be 1.5 (at $n_{H}=3.6\times10^{17} cm^{-3}$) and 1.7 (at  $n_{H}=1.2\times 10^{18} cm^{-3}$). This quantifies the tetragonal distortion of the Fermi surface. If we assume that the Fermi surface is a sphere, our measured frequency of quantum oscillations yields a carrier concentration  close to, but systematically lower than, the Hall carrier density. The two set of numbers are given in Table S2. Approximating the Fermi surface to be a squeezed ellipsoid, the magnitude of tetragonal distortion, i.e. the ratio of the two perpendicular projections is the square of the two carrier densities. As seen in Fig. S4b, our data yields a magnitude for tetragonal distortion comparable to the value measured in ref. \cite{allen}.

\begin{table*}\centering
 \renewcommand{\thetable}{S2}
 \caption{Carrier concentration and the frequency detected below n$_{c1}$.}
 \begin{ruledtabular}
\begin{tabular}{c|c|c|c|c}
Sample   & $n_1=n_H(cm^{-3})$    & $F_1 (T)$  & $n_2=n_{SdH}(Sphere)(cm^{-3})$ & $(\frac{n_1}{n_2})^{2}$   \\
\hline

SrTiO$_{3-\delta}$ & 1.58 10$^{17}$ &  8.7     & 1.45 10$^{17}$ & 1.19 \\

SrTiO$_{3-\delta}$ & 5.5 10$^{17} $ &  18.2    & 4.38  10$^{17}$  &1.58 \\

SrTiO$_{3-\delta}$ & 5.7 10$^{17} $ &  19.2    & 4.74  10$^{17}$  &1.45 \\

SrTiO$_{3-\delta}$ & 6.79 10$^{17}$ &  21.3    & 5.54 10$^{17}$   &1.50 \\

SrTiO$_{3-\delta}$ & 7.67 10$^{17}$ &  23.5    & 6.42  10$^{17}$  &1.42 \\

SrTiO$_{3-\delta}$ & 1.06 10$^{18}$ &  26.5    & 7.69 10$^{17}$   &1.90 \\

\end{tabular}
\end{ruledtabular}

\label{TabII}
\end{table*}

\end{document}